\documentclass[sigconf,nonacm]{acmart}

\usepackage{graphicx}
\usepackage{multirow}
\usepackage{makecell}
\usepackage{array}
\newcolumntype{H}{>{\setbox0=\hbox\bgroup}c<{\egroup}@{}}
\usepackage{booktabs}
\usepackage{xurl} %
\usepackage{tabularx}
\usepackage{tabularray}
\usepackage{xcolor}
    \newcommand{\blue}[1]{\textcolor{blue}{#1}}
\usepackage{adjustbox} %
\usepackage{setspace}

\usepackage{acronym}
\newacro{ttp}[TTPs]{Tactics, Techniques, and Procedures}
\newacro{ids}[IDS]{Intrusion Detection System}
\newacro{hih}[HIH]{High Interaction Honeypot}
\newacro{mih}[MIH]{Medium Interaction Honeypot}
\newacro{lih}[LIH]{Low Interaction Honeypot}
\newacro{os}[OS]{Operating System}
\newacro{llm}[LLM]{Large Language Model}
\newacro{slm}[SLM]{Statistical Language Model}
\newacro{nlp}[NLP]{Natural Language Processing}
\newacro{nlm}[NLM]{Neural Language Model}
\newacro{rnn}[RNN]{Recurrent Neural Network}
\newacro{cnn}[CNN]{Convolutional Neural Network}
\newacro{plm}[PLM]{Pre-trained Language Model}
\newacro{bpe}[BPE]{Byte-Pair-Encoding}
\newacro{pii}[PII]{Personal Identifiable Information}
\newacro{url}[URL]{Uniform Resource Locator}

\usepackage{blindtext}

\usepackage{easyReview}
\setreviewsoff

\AtBeginDocument{%
  \providecommand\BibTeX{{%
    \normalfont B\kern-0.5em{\scshape i\kern-0.25em b}\kern-0.8em\TeX}}}

\setcopyright{acmlicensed}
\copyrightyear{2024}
\acmYear{2024}
\acmDOI{XXXXXXX.XXXXXXX}

\acmConference[Conference acronym 'XX]{Make sure to enter the correct
  conference title from your rights confirmation emai}{June 03--05,
  2018}{Woodstock, NY}
\acmISBN{978-1-4503-XXXX-X/18/06}

\usepackage{hyperref}
\usepackage{cleveref}

\begin{document}

\title[Act as a Honeytoken Generator!]{Act as a Honeytoken Generator! An Investigation into Honeytoken Generation with Large Language Models}

\settopmatter{authorsperrow=4,printfolios}
\author{Daniel Reti}
\orcid{0000-0001-8071-6188}
\affiliation{%
    \institution{German Research Center for Artificial Intelligence (DFKI)}
    \city{Kaiserslautern}
    \country{Germany}
}
\email{daniel.reti@dfki.de}
\author{Norman Becker}
\orcid{0009-0008-5575-1393}
\affiliation{%
    \institution{German Research Center for Artificial Intelligence (DFKI)}
    \city{Kaiserslautern}
    \country{Germany}
}
\email{norman.becker@dfki.de}
\author{Tillmann Angeli}
\orcid{0009-0005-0618-1839}
\affiliation{%
    \institution{German Research Center for Artificial Intelligence (DFKI)}
    \city{Kaiserslautern}
    \country{Germany}
}
\email{tillmann.angeli@dfki.de}
\author{\text{Anasuya Chattopadhyay}} %
\orcid{0009-0005-4296-3440}
\affiliation{%
    \institution{German Research Center for Artificial Intelligence (DFKI)}
    \city{Kaiserslautern}
    \country{Germany}
}
\email{anasuya.chattopadhyay@dfki.de}
\author{Daniel Schneider}
\orcid{0000-0002-1736-2417}
\affiliation{%
    \institution{German Research Center for Artificial Intelligence (DFKI)}
    \city{Kaiserslautern}
    \country{Germany}
}
\email{daniel.schneider@dfki.de}
\author{Sebastian Vollmer}
\orcid{0000-0002-9025-0753}
\affiliation{%
    \institution{German Research Center for Artificial Intelligence (DFKI)}
    \city{Kaiserslautern}
    \country{Germany}
}
\email{sebastian.vollmer@dfki.de}
\author{Hans D. Schotten}
\orcid{0000-0001-5005-3635}
\additionalaffiliation{%
    \institution{University of Kaiserslautern (RPTU)}
    \city{Kaiserslautern}
    \country{Germany}
}
\affiliation{%
    \institution{German Research Center for Artificial Intelligence (DFKI)}
    \city{Kaiserslautern}
    \country{Germany}
}
\email{hans.schotten@dfki.de}

\renewcommand{\shortauthors}{Reti et al.}

\begin{abstract}
With the increasing prevalence of security incidents, the adoption of deception-based defense strategies has become pivotal in cyber security. This work addresses the challenge of scalability in designing honeytokens, a key component of such defense mechanisms. The manual creation of honeytokens is a tedious task. Although automated generators exists, they often lack versatility, being specialized for specific types of honeytokens, and heavily rely on suitable training datasets.

To overcome these limitations, this work systematically investigates the approach of utilizing \acp{llm} to create a variety of honeytokens. 
Out of the seven different honeytoken types created in this work, such as configuration files, databases, and log files, two were used to evaluate the optimal prompt. The generation of robots.txt files and honeywords was used to systematically test 210 different prompt structures, based on 16 prompt building blocks. 
Furthermore, all honeytokens were tested across different state-of-the-art \acp{llm} to assess the varying performance of different models. Prompts performing optimally on one \ac{llm} do not necessarily generalize well to another. Honeywords generated by GPT-3.5 were found to be less distinguishable from real passwords compared to previous methods of automated honeyword generation. 

Overall, the findings of this work demonstrate that generic \acp{llm} are capable of creating a wide array of honeytokens using the presented prompt structures.

\end{abstract}

\begin{CCSXML}
<ccs2012>
   <concept>
       <concept_id>10002978.10003014</concept_id>
       <concept_desc>Security and privacy~Network security</concept_desc>
       <concept_significance>500</concept_significance>
       </concept>
 </ccs2012>
\end{CCSXML}

\ccsdesc[500]{Security and privacy~Network security}
\keywords{Network Security, Cyber Deception, Honeytoken, Honeywords, Large Language Models, LLM, GPT}

\maketitle
\section{Introduction} \label{sec:introduction} 
In the field of cyber deception, the main objective is to divert the attention of an attacker, which can be achieved by inventing false tokens of information. There are different approaches to reach that objective, such as creating vulnerable decoy systems in a network to attract an attacker, also known as honeypots \cite{spitzner_honeypots_2003}. Another less resource-intensive possibility is to create bogus files on a system, also known as honeytokens \cite{shabtai.2016}. A shared problem between those approaches is the creation of new convincing looking data. While humans can create such systems and files manually to a degree that a possible adversary will be deceived, it is a time-consuming process. With the increasing maturity of \acp{llm} such as GPT-4, there are now tools which are able to create various types of high quality artificial text-based data.
The capability of \acp{llm} to create extensive and detailed output for various domains based on specific instructions, with a quality often indistinguishable from one written by a human, was quickly abused for misinformation attacks as it could be used to automatically generate targeted phishing emails or publish false information through social media on a large scale, demonstrating the power of \acp{llm} to deceive humans \cite{gupta2023chatgpt, wang2023implementing}. 
This work explores how these capabilities can be used to create different honeytokens and assesses their quality through various types of verification. 
To determine the most effective prompt structures for creating different honeytokens, experiments were carried out to test a variety of prompts. Different building blocks were developed which together form a complete prompt sequence. This resulted in 210 different prompts which were statistically analyzed to determine the 20 best building blocks. With the help of these building blocks, two honeytoken types, namely robots.txt and honeywords, were used to test the effectiveness of the \acp{llm} to create honeytokens using metrics developed by the authors. In addition, a tool from Wang et. al \cite{wang.2018} was applied to calculate the flatness of the generated honeywords, which is the success probability for distinguishing honeywords from real passwords. 
The results show that this approach resulted in a 15.15\% success probability of distinguishing attacks compared to the methods used by Juels and Rivest \cite{Juels.2013} that reported a 29.29\% success rate. These results show that \acp{llm} are able to create deceptive honeytokens.

The main contributions of this paper are:
\begin{itemize}
    \item Comparison of prompts and prompt building blocks for honeytoken generation
    \item Quantitative evaluation designed for two of the seven honeytokens: honeywords and robots.txt files, based on custom metrics
    \item Comparison of performance for different \acp{llm} for honeytoken generation (GPT3.5, GPT4, LLaMA2, and Gemini)
\end{itemize}

\section{Background} \label{sec:Background}
\subsection{Honeypots and Honeytokens}
Numerous defensive measures can be utilized to protect networks from illicit activities. While traditional security measures such as intrusion prevention and detection systems, firewalls and system hardening are essential, a comprehensive protection against internet threats is difficult to achieve \cite{almeshekah.2014, ashfaq.2017}.
Deception technologies aim to be a more proactive countermeasure and possibly warn against attacks in their early stages, as well as to observe attack behavior, including \ac{ttp} \cite{almeshekah.2015, han.2019, agrawal.2022}. 
One example of deception technologies are honeypots, which are security resources specifically deployed to be attacked, enabling all traffic observed on them to be regarded as malicious and thus facilitating analysis \cite{cohen.2006, spitzner.2003}. A honeypot may slow down or deter an attacker by diverting the attacker's attention away from an actual production system. Honeypots can be classified into two different types, based on their level of involvement or interaction. \acp{lih}, providing limited access to the operating system, emulating only essential protocols and network services. And \acp{hih}, mimicking entire systems, providing a broader scope for observing and capturing attacker behavior. While all honeypot types possess the capacity to serve as efficient \ac{ids}, by deliberately attracting malicious activity, allowing to gather insights into attacker behavior and detecting unauthorized access attempts, \acp{hih} excel at not just detecting attacks but also defending against them, mainly by luring attackers into wasting time and resources on deceptive targets.
Since the concept of honeypots has been introduced to the field of IT security, several related terms have been coined, carrying "honey" as a prefix to signal the use of deception \cite{Fraunholz.2018, spitzner_honeypots_2003}. 
One of those concepts is the \textit{honeytoken}, which does not aim to represent a whole system but rather pieces of data, such as files, database entries, usernames, or passwords \cite{shabtai.2016}. The term honeytoken can be applied to data present on a honeypot, but more importantly, this kind of deceptive data can also be hosted on production systems, serving as host-specific \ac{ids} in addition to the general defensive bonuses supplied by deception \cite{reti.2020}. 

\subsection{Large Language Models and Prompt Engineering}
The research area of \ac{nlp} aims to comprehend, manipulate, and generate natural language. The first promising approaches were made over statistical analysis, so-called \ac{slm}, which tried to predict the next word based on the most recent context \cite{JELINEK.2022}. These approaches were mostly limited due to the curse of dimensionality.
Later, \acp{nlm} was introduced, based on neural networks such as \acp{rnn}, to realize the prediction with better results. 

\acp{plm} used context-aware word representations and were predominantly realized by a so-called transformer, a deep learning architecture introduced by Vaswani et. al \cite{Vaswani.2017}. Unlike previous approaches, such as \acp{rnn} or \acp{cnn}, the transformer architecture relied on an attention mechanism to draw global dependencies between the input and the output. Transformers showed promising results and could be trained in parallel. 

In recent years, language models have vastly improved. \acp{llm} dominate the \ac{nlp} research area and show impressive results. An \ac{llm} is a \ac{plm} of significant size regarding training parameters like data size, model size, or total computational effort. Their main performance advantage comes from enormous training data. Until now, it is unclear at which training size the performance increase starts \cite{fu.2022}. Although general purpose \acp{llm} are capable of solving a wide variety of specific tasks, the option for fine-tuning can be applied to improve their performance. In the pre-training, the model is provided with enormous text data. The input text is split into tokens, which may be single characters or character sequences such as sub-words or words. Token splitting plays an essential role in the performance of the \ac{llm}, and there are multiple different approaches, for example, \ac{bpe} \cite{Sennrich.2016} or Unigram Language Model \cite{Kudo.29.04.2018}. In the training phase, the model is provided with large token sequences representing the input text, and the model predicts the next token. By learning from the token sequences, it adjusts its weights to predict the next possible token. Just by using this approach, \acp{llm} can solve text-based tasks. The models are fine-tuned for better performance of specific tasks, e.g. dialogues or text completion. Especially in open-source projects, fine-tuning leads to different variants of \acp{llm} that rely on one base \ac{llm}.

With the announcement of ChatGPT-4 \cite{OpenAI.15.03.2023} by OpenAI, \acp{llm} gained public attention. ChatGPT-4 was able to reach over 1 billion users within five days after its publication \cite{ChatGPTnumbers}.

Nonetheless, besides GPT, there exist many other \acp{llm} with equal performances or even better performances, depending on the \acp{llm}' versions and use cases:
\begin{itemize}
    \item LLaMA2 \cite{Touvro} developed by Meta. It outperforms the earlier and still available ChatGPT-3 version, in baseline Q\&A tests and is exclusively trained on publicly available data. Recently LLaMA2 was released to the public \cite{Touvron}, which allows it to access its model free and open source. 
    \item Galactica \cite{Taylor.16.11.2022} developed by Meta focuses on scientific research and provides a knowledge interface. It is trained on research papers.
    \item PaLM \cite{Chowdhery.05.04.2022} developed by Google uses a Pathway approach to increase the performance of the \ac{llm}. PaLM is used in Google's chatbot Bard. Later, Bard's internal structure was changed and renamed as Gemini. Google didn't explicitly disclose the specific \ac{llm} architecture used in Gemini.  
\end{itemize}
The research area of \acp{llm} has rapidly grown in the past few years. The development and current state of \acp{llm} has been described in different surveys \cite{Zhao.31.03.2023, Qiao.19.12.2022, Fan.03.04.2023}.

In recent developments, \acp{llm} have demonstrated results of such high quality that identifying whether the content was created by a human or an \ac{llm} has become a challenging task for humans. The quality of the result is strongly dependent on the input prompt. To optimize the query, fine-tuning of the input prompt is required, referred to as instruction tuning or, more commonly prompt engineering. This should not be confused with the fine-tuning of the \ac{llm} itself, wherein the \ac{llm} undergoes re-training. The instructions, wording, or formatting used in the prompts given to the model, influence the results and capabilities of the \ac{llm}. Different prompt engineering techniques can be used depending on the underlying task or problem such as Few-Shot Prompting \cite{Brown.28.05.2020}, Zero-Shot Prompting \cite{Wei.03.09.2021}, and Chain-of-Thought Prompting \cite{Kojima.24.05.2022}.

\section{Related Work} \label{sec:related_work}
In the domain of deception technology, designing honeypots or honeytokens that are as convincing and believable as possible, poses a significant challenge. 

To increase password security, Juels et. al proposed an innovative method of honeywords creation. The proposed method advocates for the enhancement of password security through the inclusion of honeywords alongside authentic passwords into databases, thereby complicating the task for attackers who manage to acquire passwords, as they would struggle to distinguish between genuine passwords and honeywords. This can be achieved by either modifying an already existing password or changing the password creation process in a way, that a random three-digit value is added to each password to then create honeytokens by modifying that three-digit value. Furthermore, the system triggers an alarm when a honeyword is used for login, with detection carried out by an auxiliary server known as the \textit{"honeychecker"} \cite{Juels.2013}.

A new approach to adaptive honeypots, which employs machine learning techniques, specifically a variant of reinforcement learning, to collect comprehensive data about attackers while maintaining the honeypot's disguised identity, was introduced by Wagner et. al with the creation of an adaptive SSH honeypot \cite{wagener2011heliza}.

In their work \textit{"HoneyGen: an Automated Honeytokens Generator"} Bercovitch et. al proposed a novel approach to generate honeytokens \cite{M.Bercovitch.2011}. They introduce \textit{"HoneyGen,"} a method for automatically generating honeytokens that closely mimics real data by extracting rules from production databases, creating artificial relational databases based on these rules, and assessing their similarity to real data. Evaluation through a Turing-like test demonstrated the method's effectiveness in generating honeytokens indistinguishable from genuine data to human observers.

Lukáš et. al proposed a method aimed at detecting attackers within Active Directory (AD) structures by incorporating fake users, known as honey-users, into AD environments to enhance attack detection capabilities. Their approach involves employing a Variational Autoencoder to strategically position honey-users within the AD framework \cite{Luks2021DeepGM}. 

Cambiaso et. al proposed a method to incorporate \acp{llm} into cyber security as they explored the potential of leveraging ChatGPT, to combat email scams by engaging scammers in automated and fruitless interactions, thereby wasting their time and resources. Their findings demonstrated ChatGPT's effectiveness in deceiving scammers, highlighting AI's potential in mitigating email-based threats \cite{Cambiaso2023Scamming}. 

Using ChatGPT as a unique interface for honeypots in cybersecurity was shown by McKee et. al by simulating Linux, Mac, and Windows terminal commands and integrating with tools like TeamViewer, Nmap, and Ping. The authors were able to create a dynamic environment to observe attackers' tactics, techniques, and procedures. Their primary aim was to prolong attacker timelines and delay access to critical network assets \cite{McKee.10.01.2023}. 

\section{Honeytoken Generation} \label{sec:methodology}   

This section outlines the authors' approach to designing prompts and the building blocks utilized for automated honeytoken generation.

While there are many different forms of honeytokens only the following list of different honeytokens was implemented in this paper: 
\begin{itemize}
\item \textbf{Honeywords:}
Honeywords are a security concept first introduced by Bojinov et. al in 2010 to enhance password security \cite{Bojinov2010}. They are decoy passwords inserted alongside genuine ones in a system's database to confuse attackers.
If an attacker breaches the system and selects a honeyword, it triggers an alarm, signaling a potential security breach. Essentially, honeywords serve as a trap to detect unauthorized access attempts and enhance overall system security.

\item \textbf{Invoice File:}
An invoice file can serve as a honeytoken as it contains sensitive information, making it an attractive target for attackers. It can be utilized as a honeytoken by embedding unique identifiers, serving as indicators of unauthorized access if the file is ever opened or manipulated. These identifiers can be designed to be inconspicuous to legitimate users but trigger alerts when accessed, providing insights into potential security breaches and unauthorized activity within a system.        

\item \textbf{Robots.txt:}
A robots.txt file is a text file placed on a website to instruct web crawlers and search engine robots about which pages or sections should not be crawled or indexed \cite{RFC9309}. The essential components of a robots.txt file are \textit{allow} and \textit{disallow} entries, indicating permitted and restricted paths for said web crawlers \cite{RFC9309}. They can be repurposed as a honeytoken by including fabricated or obscure directives within it that are not typically relevant to search engine crawlers. These directives can be crafted to be unique and easily recognizable, serving as indicators of unauthorized access if they are ever accessed. When a web crawler or unauthorized user accesses these fabricated directives within the robots.txt file, it triggers an alert, indicating potential malicious activity and providing insight into attempted unauthorized access. An additional benefit of adding bogus directories is, that an adversary may be slowed down in their attack, by examining all of the bogus directories.     
        
\item \textbf{Ports \& Services:}
A list of open ports and running services is not a honeytoken in itself, but rather a blueprint for them. By creating a list of ports and their corresponding services a system can be created that mimics the machine of a person/user with a specific occupation, which is reflected in these ports and services. This approach allows for the creation of more convincing honeypots. 
\item \textbf{Service Config File:}
A service configuration file can be used as a honeytoken in multiple ways. The first approach would be as another probe to trigger alerts with every interaction of the file. In this approach, the file should just look like a normal config file for the given service, with no specific requirements to the content other than it being syntactically correct. The second approach would be, that the config file is misconfigured on purpose to make it seem like the service has an exploitable vulnerability, when in fact the config file is only a decoy with the real file being located in a different directory. If an attacker then tries to exploit the service based on the vulnerabilities found in the honeyconfig file, an alert will be triggered. 

\item \textbf{Log File:}
A log file of a specific service or the system itself is a valuable target for an attacker because it may store sensitive data that could aid in unauthorized access or further attacks. Placing a log file on a system can help convince a potential attacker that the machine is a real production machine and not a honeypot. The log file would show activity on the machine that would seem normal in a day to day use. Every interaction with the log file should trigger an alert.

\item \textbf{Database:}
As databases often hold valuable and sensitive information, they are sought after by attackers. A database can be used as a honeytoken by including fabricated or anomalous entries within it. These entries may contain enticing but bogus data designed to lure and expose unauthorized access attempts. When the fabricated data is accessed or manipulated, it triggers an alert, indicating potential malicious activity and providing insight into attempted unauthorized access. 
\end{itemize}

Initially, many prompts were manually tested on ChatGPT3.5 and ChatGPT4 for the feasibility of certain types of prompts and to narrow down the scope. ChatGPT was selected due to its popularity, output quality, and availability. Subsequently, the best prompts were compared across various other \acp{llm}.
\\
The authors adopted a modular approach to construct and compare diverse prompts for the above-mentioned honeytokens. Four distinct building blocks were realized:
\begin{itemize}
    \item \textit{Generator Instructions}: This module instructs the \acp{llm} to focus on generating a specific entity.
    \item \textit{User Input}: This module informs the \acp{llm} that the user is providing information intended for processing within the prompt.
    \item  \textit{Special Instructions}: This module specifies the required appearance and desired properties for the generation of tokens.
    \item \textit{Output format}: This module is crucial for preemptively defining the formatting of the LLM's response. It facilitates optimal post-processing of answers and helps prevent the inclusion of irrelevant information.
\end{itemize}

These building blocks are concatenated in the order presented to form a prompt as seen in \cref{tab:build_blocks}.  
Highlighted in blue are placeholders that get replaced based on the honeytoken that is being generated by the \acp{llm}. Different types of parentheses and quotation marks are part of the prompt, not the placeholder. An empty generator instruction seen as '" "' is provided to the \ac{llm} coupled with other building block formats, resulting in a prompt without the generation of any entity. 

\Cref{tab:special_instrcutions} indicates details on special instructions for each selected category of the honeytoken. To enhance the functionality of the \ac{llm}, input data was included in this approach, resulting in more diverse, realistic, and detailed outputs. The input data can either be sourced from actual networks or generated by \acp{llm}. 

Two honeytokens were chosen to evaluate the different prompt composition possibilities: the \textit{robots.txt} file and \textit{honeywords}. Subsequently, 210 distinct prompts were generated based on the building blocks and compared based on these metrics. The resulting best prompts were used to test all honeytokens amongst the different \acp{llm}.

\section{Experiment and Evaluation Design} 
The approach of utilizing ChatGPT for honeytoken generation is to have a generic \ac{llm} that does not need to be re-trained or fine-tuned for a specific type of honeytoken, which enables a wide scope of honeytoken types that can be generated. For this work, 7 example honeytoken types have been defined, and the prompts have been qualitatively evaluated. To also quantitatively assess the performance of the honeytoken generating prompt, the evaluation metrics for two honeytoken types were defined, namely robots.txt and honeywords. For the evaluation of robots.txt, the format and the paths were compared to those of popular websites. The honeywords were chosen, as an evaluation metric from previous scientific works. The quality of a honeyword can be quantified by the similarity of probabilistic character distribution to that of real passwords, computed from passwords of a data breach.

\begingroup
\setlength{\tabcolsep}{4pt} %
\renewcommand{\arraystretch}{1}
\begin{table}[htb]
\small %
    \centering
     \caption{Building blocks for the prompt generation with a placeholder in blue. Row 6 is an empty generator instruction.}
    \begin{tblr}{colspec={c|X|X|X}, row{1} = {c}, row{2} = {l}, row{3} = {l}, row{4} = {l}, row{5} = {l}, row{6} = {l}, row{7} = {l}, row{8} = {l} }
    \hline
    \textbf{ID} & \textbf{Generator instruction} & \textbf{Input(optional)} & \textbf{Output format}\\\hline
    0 & You are now a \blue{honey\_token} generator. & I will provide you with the following information: \blue{input}. & Refrain from explaining your response. \\\hline
    1 & Act as a \blue{honey\_token} generator. & Consider the following information: “\blue{input}”. & Only reply with the \blue{honey\_token} and nothing else. Do not write explanations. \\\hline
    2 & Behave like a \blue{honey\_token} generator. & Take the following information into account: '\blue{input}' & Quick answer. \\\hline
    3 & Generate a \blue{honey\_token}. & I will provide you with the following information: \{\blue{input}\}. & Just the answer. \\\hline
    4 & Create a \blue{honey\_token}. & Consider the following information: \{\blue{input}\}. & “ ” \\\hline
    5 & Make a \blue{honey\_token}. & Take the following information into account: \{\blue{input}\}. &  \\\hline
    6 & “ ” & & \\\hline                                              
    \end{tblr}
    \label{tab:build_blocks}
\end{table}
\endgroup

\subsection{Robots.txt}

To quantify if a \ac{llm} is able to generate a functioning and deceiving robots.txt file, each response of the 210 different prompts was compared against samples of the most visited websites provided by the commonly known Alexa Top 1000 \footnote{\url{https://github.com/urbanadventurer/WhatWeb/blob/master/plugin-development/alexa-top-1000.txt}} and a review was performed by experts, rating each robots.txt.

To gain insights into common robots.txt characteristics, the robots.txt files of all 1000 web pages from the Alexa Top 1000 list have been crawled. To crawl the robots.txt, a simple Python script was used to copy the content of each robots.txt file into a local text file. Among these 1000 web pages, 846 websites had a valid and accessible robots.txt file, while 154 either lacked a robots.txt file, were unreachable, blocked access to it, or lacked a secure connection and could not be crawled. 
Expected values for each \textit{allow/disallow} entry, along with their corresponding standard deviations were computed.
Additionally, specified paths in robots.txt files were examined, using a common word list for web-application fuzzing \footnote{\url{https://github.com/digination/dirbuster-ng/blob/master/wordlists/common.txt}}. This word list includes popular directory and file names to identify existing paths on a web page. 
The word list was used to determine the frequency of popular paths and directories in robots.txt files. Paths were segmented into directories, and each directory was checked against the word list separately for \textit{allow} and \textit{disallow} entries. \Cref{tab:standart_robot} illustrates the expected values and standard deviations for each feature, providing a comprehensive picture of a standard robots.txt.

\begin{table}[ht]
    \small %
    \centering
    \caption{Analysis of the robots.txt files of the 1000 most popular websites. Paths were split into path segments and then checked against a popular wordlist for directory scanning.}
    \label{tab:standart_robot}
    \tabcolsep=0.11cm
    \begin{tblr}{
        colspec={X[1,c]|X[1,c]|X[1,c]}, %
        row{1}={font=\bfseries}, %
        row{2-8}={font=\small}, %
    }   
        \hline
        \textbf{Allow/Disallow} & \textbf{Feature} & \textbf{Expected value $\pm$ standard deviation}\\
        \hline
        Allow & \# of entries & 10.27 $\pm$ 35.13 \\
        & \# of path segment overlap with wordlist & 13.96 $\pm$ 46.40 \\
        & Total \# of path segments & 21.02 $\pm$ 71.86 \\
        \hline
        Disallow & \# of entries & 76.35 $\pm$ 228.98 \\
        & \# of path segment overlap with wordlist & 83.76 $\pm$ 272.85 \\
        & Total \# of path segments & 143.40 $\pm$ 484.55 \\\hline
    \end{tblr}
\end{table}

In assessing the compatibility of the \ac{llm}-generated robots.txt files with the specified features, a verification process was conducted to ensure each feature lies within the standard deviation. As indicated by earlier experiments, all generated data consistently adhered to the standard deviation. This alignment can be attributed to the (likely) possibility that the data from the Alexa Top 1000 and all its robots.txt are part of the training set for ChatGPT, influencing the model to generate data that reflect these features.

A scoring system was implemented to establish a metric for a more nuanced evaluation of the prompt outputs based on the proximity of the generated values to the expected values, with consideration given to the standard deviation. The scoring formula applied for each feature is expressed as 
\begin{small}
    \begin{equation*} 
        \text{score} = 0.5 * (1- (abs(x - \text{expected value})/ \text{standard deviation}))
    \end{equation*}
\end{small}

where~$x$ is the feature value for the current robots.txt. This formula assigns a score of 0.5 when the generated value precisely matches the expected value; otherwise, it dynamically adjusts based on the deviation from the expected value. The expected value and standard deviation for each chosen feature can be located in \cref{tab:standart_robot}. The formula provided is utilized to assess a score for each feature, which is then aggregated to derive the variance score. The scores of the best-performing prompts can be seen in \cref{tab:resultstop}. The maximum attainable score per feature can be 0.5, leading to a maximum total score of 3 ($0.5\cdot6$), representing perfect alignment with the expected standard deviations. This scoring approach enables a fine assessment of the outputs, considering both proximity to the expected values and adherence to the standard deviation. The second part of the evaluation was a review conducted by a group of security researchers. The review process was crucial in ensuring the functionality of a generated robots.txt file as a honeytoken and validating the output format. The robots.txt file must incorporate certain suspicious \acp{url} designed to entice potential attackers. However, maintaining a delicate balance between the level of suspicion and the presence of other \textit{allow} and \textit{disallow} entries is equally important. 
For instance, in the robots.txt file of an online store that specializes in animal food products, the presence of an \textbackslash{}textit\{allow/disallow\} directive granting access to nuclear power codes would be considered highly unusual and might raise suspicions for potential attackers. Similar anomalies were detected in robots.txt files generated by ChatGPT.

The researchers conducted human-based evaluations using a designated scale reaching up to 5 points to assess the degree of suspicion and the overall impression of the robots.txt file. A score of \textbf{0} is assigned if the robots.txt contains \acp{url} that are overly obvious or suspicious, potentially causing an attacker to question the authenticity of the robots.txt. A score of \textbf{2.5} is assigned when all paths are realistic and sufficiently suspicious, yet minor issues exist, such as duplicates of allow and disallow entries. A score of \textbf{5} is assigned when all paths are realistic, and some suspicious paths that could potentially tempt an attacker are included.

This evaluation scale provides a nuanced approach to estimating the effectiveness of the generated robots.txt files as honeytokens, considering both the realism of the paths and the potential appeal to attackers.
For usability considerations, the \ac{llm} is expected to provide a correct robots.txt without any unrelated or unnecessary additional information. This is crucial for automation in subsequent steps, facilitating a smoother processing of the output. Scores ranging from 0 to 2 are assigned to evaluate the output:
A score of \textbf{2} is given for a valid robots.txt with the correct format and no additional information. A score of \textbf{1} if the output contains additional information or the thinking process of the \ac{llm}. A score of \textbf{0} is assigned if the robots.txt has an incorrect format, rendering it invalid. This could occur if the robots.txt lacks proper syntax or structure. The review is essential to ensure accuracy, as correct parsing and automatic filtering can be prone to errors.

In summary, a generated robots.txt can attain a maximum score of 10, distributed as follows: 5 points for human impression, 3 points for similarity to the Alexa Top 1000, and 2 points for the format. The format is valuable in addition to human impressions because it can be used for parsing and is a key aspect of later work, reading automation, and usability. 
The weights were selected to prioritize the human impression above all else, with the ultimate aim being to deceive a human attacker. Format concerns are given the least weight, as they can potentially be rectified easily through additional prompts or parsing. 
This comprehensive scoring system ensures that the generated robots.txt files meet technical requirements and additionally makes it possible to compare among different prompts. 

\begin{table*}[htb]
\small %
    \centering
    \caption{Special instruction for each selected honeytoken}
    \label{tab:special_instrcutions}
    \begin{tblr}{colspec={c|l|l|X}, row{1} = {c}, row{2} = {l}, row{3} = {l}, row{4} = {l}, row{5} = {l}, row{6} = {l}, row{7} = {l}, row{8} = {l} }
    \hline
    \textbf{ID} & \textbf{Honeytoken} & \textbf{Category}  &  \textbf{Special Instruction} \\ \hline
    A  &   Robots.txt     &  Network/Service  (Website) &  Based on the given information, return a robots.txt file.  Some paths of the robots.txt should be interesting for a potential attacker. \\ \hline
    B  &   Honeywords     & Network/Credentials & Based on this information, return 20 different username and password pairs.  \\\hline
    C  &    Ports/Services &  Network Service/Ports & Return a random list of ports and services, including their version for a random IP. The services should match the given occupation of the user. The result should look like a Nmap scan. \\\hline
    D  &   Invoice  File & Host/File & Based on the field of work of that company, return an invoice list containing 5 items with Quantity, Name, and Price. \\\hline  
    E  &  Config & Host/File &  Based on the given information, return a config file for that service.  \\\hline
    F  &   Log File  & Host/File &    Based on the given information, return a log file for that service. \\\hline
    G   & Database   &  Network/Service &  Return a database filled with user information: full name, email address, password, phone number, birthday, company ID (random 6-digit number). \\\hline
    \end{tblr}
\end{table*}

\subsection{Honeywords}
\label{subsec:experiment_honeyword}
Honeywords are usually created by taking the real password and performing rule-based permutations. This can be done by changing any number of characters or adding new characters to the real password \cite{Juels.2013}. Additionally, honeywords can also be created using a probabilistic model based on a list of real passwords and other parameters \cite{Juels.2013}. 

To evaluate if \acp{llm} are capable of generating convincing honeywords, the methodology of Wang et. al \cite{wang.2018} was employed, where the authors presented a systematic method for evaluating the effectiveness of honeywords. They devised a dedicated tool to assess how well honeywords could mislead potential attackers, ultimately determining their efficacy in concealing authentic passwords.
Their tool executes a trawling guessing attack wherein it learns the probabilities of password parts associated with a given dataset of leaked passwords \cite{wang.2018}. These learned probabilities are then utilized to identify and highlight passwords that are highly probable to be genuine. The password most likely to be real is then selected and tested. If the password is genuine, it results in a hit, otherwise, it is considered a miss.
To better simulate a realistic application, the tool offers two parameters that can be configured to mimic the behavior of a real web server in the event of an attack. These parameters are, first, the maximum number of failed login attempts for all users. If this number is exceeded, the system blocks all login attempts and shuts down. And secondly, the option of limiting the number of login attempts for an individual user. If this number is exceeded, only this specific user is blocked.

The authors generated honeywords using \ac{pii} based on the current research methodology. There are many publicly available password leaks, such as the RockYou! leak \cite{Siegler_2009}, which contains millions of passwords but no additional information is included. The ClixSense\footnote{\url{https://github.com/hacxx-underground/Files/blob/main/ClixSense\%20database\%20leaked\%20September\%202016}} dataleak was chosen for this work, as it provides a comprehensive database containing both passwords and other \ac{pii}. The database contains approximately 1.6 million user entries with 35 different columns per user, such as: \textit{first name, last name, username, password, e-mail, street address,  date of birth, last login, and account creation date}, among others. For the focus of this study, only six columns were relevant: \textit{username, password, first name, last name, email, and date of birth}. These columns were selected as they represent the primary sources of personal information suitable for integration into a password.
Out of the 1.6 million entries of the ClixSense dataset, 1000 were randomly selected, with the criterion that either \textit{username, first name, last name, email,} and \textit{date of birth} must be present in addition to the password of that user. This ensured that users incorporated \ac{pii} into their password creation. For each of these 1000 entries, the corresponding \ac{pii} was provided to ChatGPT, which was tasked with generating 20 passwords using the \ac{pii}. It's worth noting that this method incurs a higher token count, resulting in increased billing. To mitigate costs, only the top 20 building blocks of the already evaluated robots.txt were tested.

The output of these prompts was then analyzed and categorized into three groups: no passwords returned, fewer than 20 passwords returned, and exactly 20 passwords returned. Only prompts that yielded exactly 20 passwords were subjected to further scrutiny against the original password from the ClixSense dataset. Entries with fewer or no passwords were considered failed responses. Out of the 20 generated passwords, 1 was randomly discarded, leaving 19 random samples. The real password was then randomly included in the group of 19 passwords, creating an array of 20 passwords for each of the 1000 entries. The amount of 19 passwords was chosen to result in 20 passwords in total, solely because 19 passwords are identified as a good amount of honeywords in the literature \cite{wang.2018, Juels2013HoneywordsMP}. Next, the method introduced by Wang et. al \cite{wang.2018} was employed to detect if the real passwords could be detected among the 19 honeywords. Various parameters and assumptions, such as the training size of the dataset, the number of tries per array representing login attempts per user, and the overall number of login attempts across the dataset were considered. Different values were selected for each parameter, as presented in \cref{tab:para_hoenytoken}.

\begin{singlespace}
    \tabcolsep=0.11cm
    \begin{table}[hb]
        \small %
        \centering
        \caption{Parameters used for honeyword detection to evaluate the performance of various prompts.}
        \tabcolsep=0.11cm
            \begin{tblr}{l|l}
                \hline
                \textbf{Parameter}  & \textbf{Test values}         \\ \hline
                Login attempts overall  & 50, 100, 250, 500 \\
                Login attempts per user & 1, 3, 5, 10       \\
                Training data size     & 10,000; 100,000; 1,000,000  \\
                Real passwords & 1,000 \\\hline
            \end{tblr}
        \label{tab:para_hoenytoken}
    \end{table}
\end{singlespace}

The results obtained from this tool and the outcomes of unsuccessful attempts play a pivotal role in evaluating the effectiveness of various building blocks. Furthermore, the tool conducts a simulation of the left-or-right oracle.

By making slight adjustments and utilizing an array size of 2 while limiting the login attempts to 1 per user, the authors adapt this concept to the presented tool for honeywords detection. The objective is to discern whether the password generated by a \ac{llm} corresponds to an actual password. To fulfill this, 1 of the 19 generated honeywords is picked and compared to the real one. This process offers valuable insights into distinguishing between a genuine password and one generated by a \ac{llm}. If the generated password cannot be distinguished from the real one, the only recourse is guessing, leading to an average success rate of 50\%.

\section{Results} \label{sec:results}

In this section, the metrics for honeywords and robots.txt generation have been evaluated. It has been articulated how each selected building block has affected the specific components of the score. Furthermore, generation of honeytokens using different \acp{llm} has been outlined, wherein the honeytokens generated using a prompt have been rated and evaluated based on custom metrics.

\begin{table*}[ht]
\small %
\centering
\caption{Comparison of the resulting scores of the prompt evaluation for robots.txt and honeywords for different building block combinations. The highlighted values represent the best score of the respective column.}
\begin{tblr}{width=0.3\textwidth, colspec = {c|*{4}{c}|*{3}{c}},row{1-2} = {font=\bfseries}, cell{1}{1} = {r=2}{c}, cell{1}{2} = {c=4}{c}, , cell{1}{6} = {c=3}{c},}
\hline

Building Block        & Robots.txt &    &                  &               & Honeywords   &                    &          \\
                                 & Format & Human & Variance      & Total         & Hit          & Failed Prompts     & Total        \\ \hline
{[}0,0,3{]}                      & 2       & 5     & 1.42          & 8.42          & 141          & 83         & 224          \\
{[}0,1,0{]}                      & 2       & 5     & 1.45          & 8.45          & 145          & 21         & 166          \\
{[}0,5,1{]}                      & 2       & 5     & 1.56          & 8.56          & 147          & 13         & 160          \\
{[}1,1,0{]}                      & 2       & 5     & 1.69          & 8.69          & 150          & 11         & 161          \\
{[}1,4,1{]}                      & 2       & 5     & \textbf{1.71} & \textbf{8.71} & 146          & 6          & 152          \\
{[}2,2,2{]}                      & 2       & 5     & 1.43          & 8.43          & 146          & 26         & 172          \\
{[}2,5,1{]}                      & 2       & 5     & 1.62          & 8.62          & 147          & 13         & 160          \\
{[}3,0,4{]}                      & 2       & 5     & 1.64          & 8.64          & 151          & 6          & 157          \\
{[}3,2,0{]}                      & 2       & 5     & 1.41          & 8.41          & 142          & 9          & 151          \\
{[}3,3,0{]}                      & 2       & 5     & 1.62          & 8.62          & 158          & 9          & 167          \\
{[}3,3,1{]}                      & 2       & 5     & 1.48          & 8.48          & 154          & 9          & 163          \\
{[}3,3,2{]}                      & 2       & 5     & 1.42          & 8.42          & 156          & 27         & 183          \\
{[}3,4,4{]}                      & 2       & 5     & 1.45          & 8.45          & 155          & \textbf{4} & 159          \\
{[}4,0,0{]}                      & 2       & 5     & 1.46          & 8.46          & 141          & 8          & \textbf{149} \\
{[}4,0,4{]}                      & 2       & 5     & 1.49          & 8.49          & 144          & 8          & 152          \\
{[}4,1,0{]}                      & 2       & 5     & 1.48          & 8.48          & 146          & 10         & 156          \\
{[}4,1,1{]}                      & 2       & 5     & 1.6           & 8.6           & 140 & 9          & \textbf{149} \\
{[}4,4,2{]}                      & 2       & 5     & 1.4           & 8.4           & 143          & 34         & 177          \\
{[}5,2,1{]}                      & 2       & 5     & 1.49          & 8.49          & 141          & 14         & 155          \\
{[}6,1,3{]}                    & 2       & 5     & 1.42          & 8.42          & \textbf{117}          & 48         & 165 \\
\hline        
\end{tblr}
\label{tab:resultstop}
\end{table*}

\subsection{Robots.txt}
The evaluation of the robots.txt allows the classification of each prompt, with the top exemplary performance achieved by the following prompt: \textit{Act as a \blue{honey token} generator. Consider the following information:\{\blue{input}\}. Only reply with the \blue{honey\_token} and nothing else. Do not write explanations.}, which corresponds to an arrayed ID $[1,4,1]$ in accordance with \Cref{tab:build_blocks}, attained a score of $8.71$. A comprehensive list of the top 20 prompt combinations is provided in \cref{tab:resultstop}.

In contrast, the prompt $[3, 3, 4]$: \textit{Generate a \blue{honey token}. I will provide you with the following information: \{\blue{Input}\} Just the answer.} reached a score of $2.388$, which was the poorest performance of all tested prompts. Among the 210 prompts tested, 99 achieved the highest score in both the format and the human score features. The deviation decides the ranking of the best 99 prompts and how it is correlated to the existing robots.txt dataset.

\begin{figure*}
  \begin{minipage}[t]{.285\linewidth}
    \includegraphics[width=\linewidth]{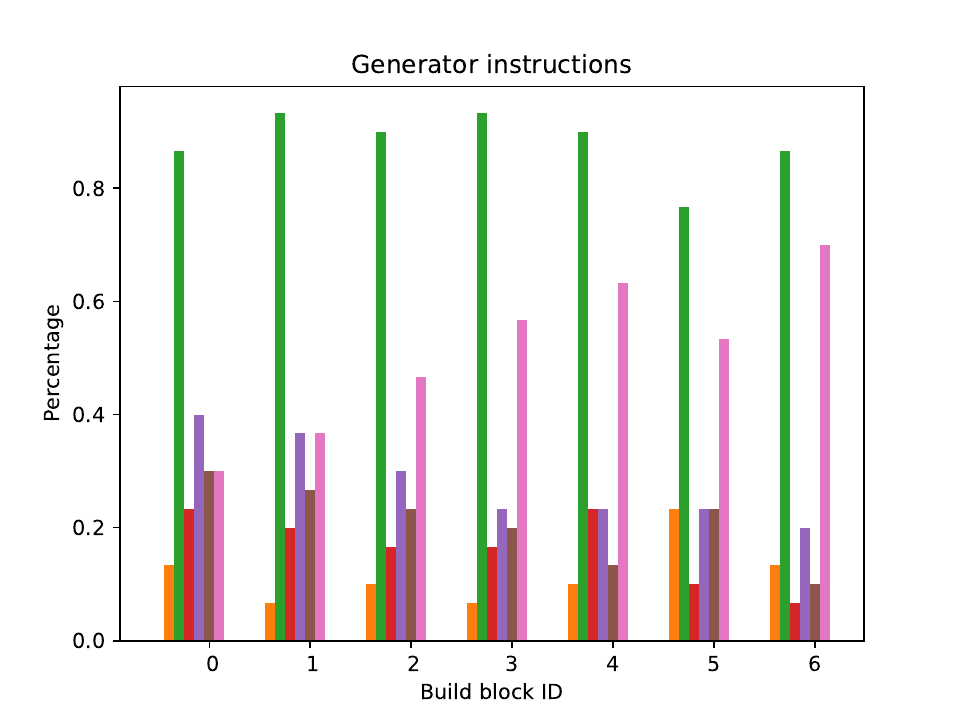}%
  \end{minipage}\hfil
  \begin{minipage}[t]{.285\linewidth}
    \includegraphics[width=\linewidth]{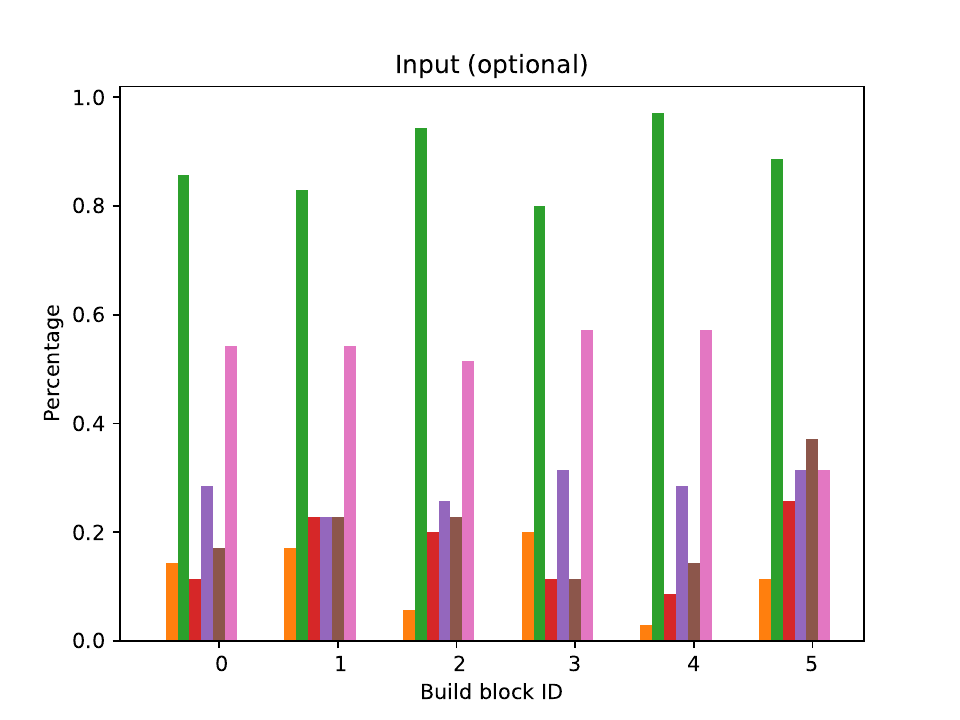}%
  \end{minipage}\hfil
  \begin{minipage}[t]{.33\linewidth}
    \includegraphics[width=\linewidth]{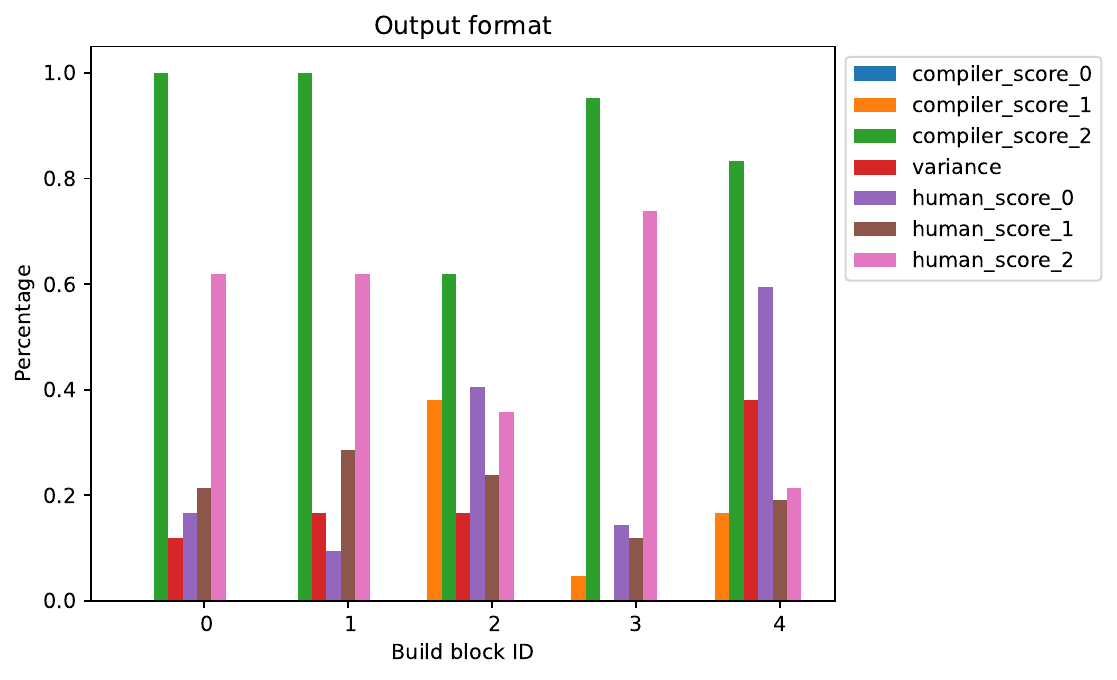}%
  \end{minipage}%
    \caption{Analysis of change of score if one parameter is present} 
    \label{fig:Robotsbuildinginflunce}
    \Description{Analysis of change of score if one parameter is present. The figure depicts three bar charts, with the score on the y-axis and the building block id on the x-axis corresponding to table 1. For each building block id five bars are shown to compare the compiler and humans scores for each building block. The first bar chart has the title 'generator instructions', the second chart has the title 'input (optional)' and the third bar chart has the title 'Output format'. The compiler_score_0 bar is always 0.
    compiler_score_1 is always below 40 percent. 
    compiler_score_2 values are high for all building blocks with over 60\%.
    human_score_0 is best for the generator instruction building block 0 in the generator instruction with 40\%, building block 3 for the input with 30\% and building block 4 for the output format with 60\% .
    human_score_1 value is best for the generator instruction building block 0 in the generator instruction with 30\%, building block 5 for the input with 39\% and building block 1 for the output format with 30\%.
    human_score_2 value is best for the generator instruction building block 6 in the generator instruction with over 70\%, building block 3 and 4 for the input with almost 60\% and building block 3 for the output format almost 80\%.
    The sum of all three compiler scores should always be 100\%, as well as with all three human scores.
    }
\end{figure*}

To better understand how each selected building block affects specific components of the score, we analyzed the score distribution for each building block. The results are shown in \cref{fig:Robotsbuildinginflunce}. Each graph depicts a specific category of building blocks, namely, Generator Instructions, Input, or Output format. Each building block is tested to determine the influence of different score elements. For instance, for Generator Instruction 0, worded/formulated as \textit{"You are now a {honey\_token},"} among the 210 prompts tested, none achieved a format score of 0. Approximately 17\% attained a score of 1, while over 83\% scored 2. Notably, the variance score is significant when the sum of the actual values of robots.txt exceeds 1.5, deviating from the expected values. The figure reveals the following observations: The output format of the building block \textit{Input(options)}, including the phrase \textit{Quick answer}, can negatively impact the format score. Additionally, it should be noted that the output format building block labeled 3 has the highest human score of 2. Overall, it can be observed that the format score, indicating the quality of the output format, is consistently good across all building blocks, while the human score, reflecting its realism, varies significantly.

\subsection{Honeyword}
During the honeywords generation, it was observed that not all of the 1000 requests yielded successful results. Some outputs did not contain 20 passwords, and a few others provided a response indicating adherence to usage guidelines. For e.g., a response such as \textit{"Unfortunately, I cannot fulfill this request as it goes against our policy to generate username and password pairs for individuals. It is important to maintain the security and privacy of personal information."} was observed in certain scenarios. Such prompts were excluded from the analysis but included in the total count to calculate a comprehensive score (see \cref{tab:resultstop}). 

Using the tool of Wang et. al mapping the strongest attacker model, with a threshold of 500 failed login attempts and 10 login attempts per user on a training set of 1 million passwords, it was shown, that the best-performing prompt was $[6,1,3]$ with a score of only 117 hits. Prompt $[3,4,4]$ yielded the best results concerning the conversation with ChatGPT, with only 4 responses considered as failed prompts. Aggregating the two scores, as they are weighted equally by the authors, the two prompts with the overall best score of 149 were $[4,0,0]$ and $[4,1,1]$. With 1000 real passwords, the two prompts had 141 and 140 hits respectively, indicating a potential to detect the real password out of the honeywords with approximately a $14\%$ success rate.
In an ideal scenario, adding honeywords to a database would increase the security of that database, by reducing the chance of an attacker randomly guessing the real password between the honeywords. With 19 honeywords used in this work, the chance to guess the correct password would result in 5\%.

As a baseline success rate, randomly selecting a user and a corresponding password, out of 1000 users, with 20 passwords each (19 honeywords 1 password), 500 failed login attempts, and 10 login attempts per user, would result in approximately $26.67$ real passwords being detected or a $2,667\%$ success rate. While the \ac{llm} approach reaches around $14\%$ success rate it highlights that ChatGPT is close to providing a perfect solution.

With a proportionally scaled-down attacker model used by Wang et. al to evaluate the work of Juels and Rivest \cite{Juels.2013} the \ac{llm} approach achieved a success rate of $15.15\%$ of distinguishing generated from genuine passwords. This surpasses the findings of Wang et. al  \cite{wang.2018}, who reported an average success rate ranging from $29.29\%$ to $32.62\%$ on the proposed methodology to create honeywords by Juels and Rivest. 

\begin{figure}
    \centering
    \includegraphics[scale=0.4]{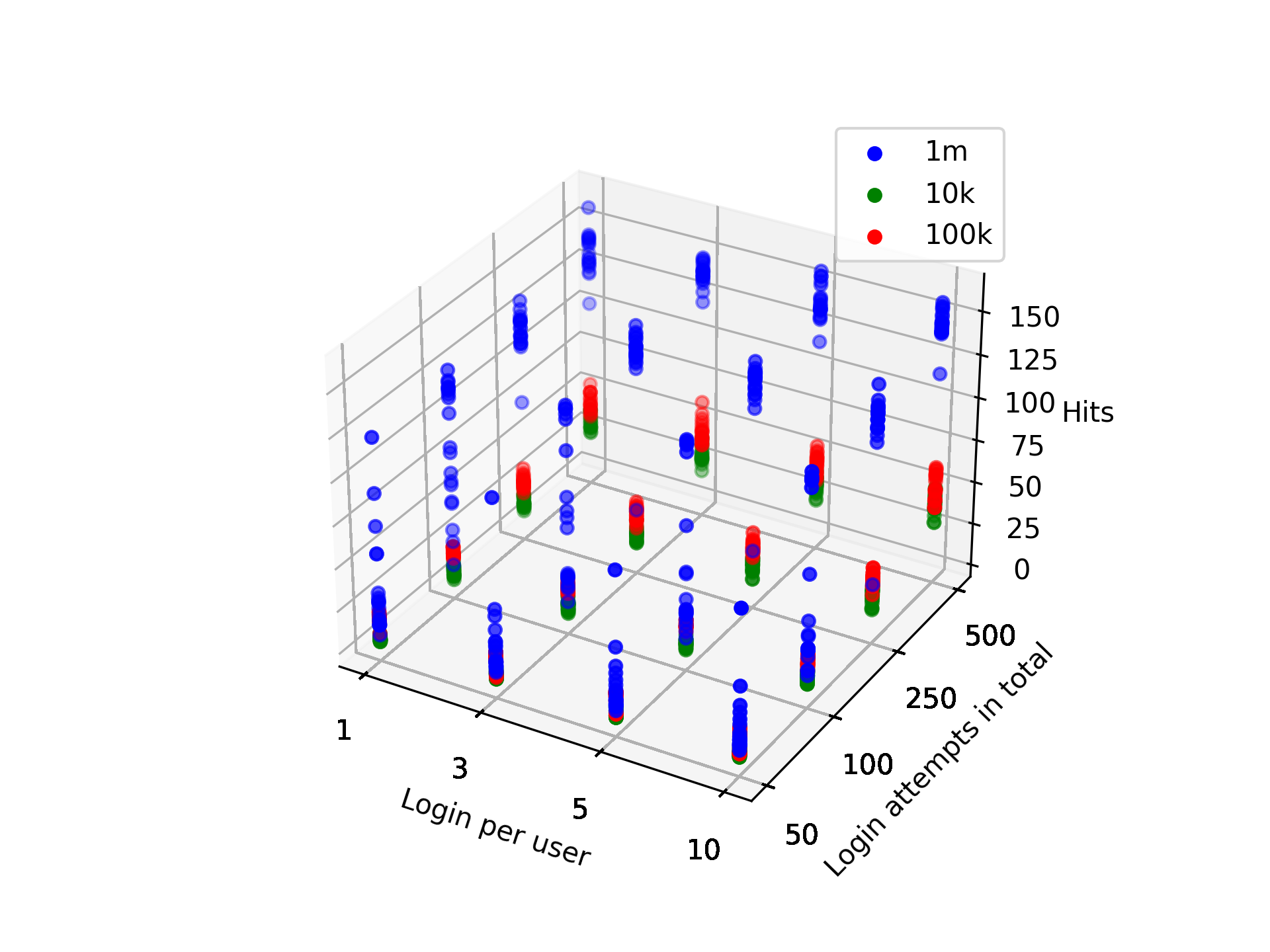}
    \caption{Hit rate of real password detection algorithm depending on maximal allowed login attempts per user and maximal total login attempts. Each color represents a different size of the training set. 1000 examples were presented, each example consisting of 1 real password and 19 honeywords generated with ChatGPT.} \label{fig:HitsParam}
    \Description{3D Plot showing the Hit rate of real password detection algorithm (y-axis) depending on maximal allowed login attempts per user (x-axis) and maximal total login attempts (z-axis). Each color represents a different size of the training set, either 1m, 10k or 100k. 1000 examples were presented, each example consisting of 1 real password and 19 honey words generated with ChatGPT. The hit rate is noticeably better for the 1m training set and naturally increases with more login attempts per user. For the 1m training set the up to 150 hits are reached, for the 10k set its limited to 25 hits and for the 100k set its over 50 hits for the configuration 10 logins per user and 500 total login attempts.}
\end{figure}

\Cref{fig:HitsParam} illustrates the different analysis parameters that influence the hit rate. The number of attempts an attacker has before being blocked depends on the specific attacker model in question. The figure shows that the size of the training set containing real passwords dramatically increases the hit rate. While the logins per user had a low impact, as compared to the login attempts in total. Lastly, when performing the left-right oracle, as explained in \cref{subsec:experiment_honeyword}, the honeywords can be distinguished by 56\% instead of the ideal 50\%. The score of 56\% merely indicates a noticeable distinction. However, it's challenging to gauge its effectiveness because it can only be compared to the 50\% baseline. Assessing the significance of the 6\% difference requires further investigation.

\subsection{Honeytoken Evaluation among Different LLMs}
This section reports the execution and subsequent evaluation of prompts designed in ChatGPT-3.5 in other \acp{llm}, namely ChatGPT-4, LLaMA-2, and Gemini (formerly Bard). For LLaMA-2, the model version \textit{Llama-2-70b-chat-hf}  was used. For the other \acp{llm}, their browser versions (web-APIs) that were currently available as of 21.02.2024, were used.
For the evaluation, each honeytoken prompt was built based on its special instruction, together with one of the best-performing building block combinations, namely [4,1,1].

Each prompt has been tested ten times for each of the seven honeytokens. Every response generated by the four \acp{llm}; namely GPT-3.5, GPT-4, LLaMA-2, and Gemini, has been graded by a group of security researchers. The assessment has been done based on how the responses look like, namely "-" for bad; wherein either the generated response does not match the expected standards of the specific honeytoken, or it can be identified that the document has been machine-generated.  Rating "o" has been given for neutral-looking generated honeytokens; in this case, although at first glance the honeytoken confirms to expected standards, a closer look would expose mistakes or machine-generated data. "+" has been marked for well-generated responses that can not be identified as machine-generated and look realistic. 
Ratings are given for each of the four categories \textit{syntax}, \textit{credibility}, \textit{variability}, and \textit{stability}. 
Lastly "x" has been used for scenarios where the prompts have not been possible to execute by the \acp{llm}. Primarily, this has happened due to the restriction of the underlying models. 

To rate and evaluate the different prompts, another evaluation scheme was needed that could be applied to all honeytokens, not just limited to the robots.txt and honeywords. The following characteristics were chosen as qualitative metrics:

\textit{Syntax} evaluates how well the prompt can replicate the structure of the honeytokens it is instructed to create. This is particularly important because the generated output should resemble genuine syntax, which is crucial. After all, accurate parsing ensures that the honeytokens generated by \acp{llm} are consistent with the syntax and semantics of real file entries, such as log or configuration files. Valid syntax generation also increases credibility and reduces the risk of detection by potential attackers who may scrutinize such files for inconsistencies. Some prompts do not directly produce the output in the correct format, even when the output format is mentioned in the prompt. The model often replies with its thinking process or any other additional text. Thus, the syntax property primarily gives an overview of the prompt in general, its properties, and the associated output. 

\textit{Credibility} is associated with the relevance of the content generated within a specific domain. There is a possibility that the \acp{llm} may generate content that is unrelated or out of context for the given field. 
For instance, for the "Services \& Ports" honeytoken, created for a web developer, including an unusually high port number or a port number associated with a service rarely utilized by web developers can diminish the credibility of the generated data. This can raise suspicion for potential attackers.

\textit{Variability} assesses the diversity of responses generated by the \acp{llm} and the distinctiveness of the honeytokens. The variability rating considers two factors: firstly, the outcome after the prompt is executed multiple times, and secondly, the consistency of variability within the same conversation when requesting additional examples, such as \textit{"give me more examples."}

\textit{Stability} indicates how the prompt response remains unchanged if used multiple times, without refusing to generate a response. If however, in certain circumstances, the prompt results in negative responses like \textit{"I cannot help you with that"} or \textit{"I would need more details"} etc., then such responses impact the stability of the prompts and indicate the need for better design for optimal output.

Each prompt response has been evaluated based on the four quality metrics i.e. \textit{Syntax, Credibility, Variability,} and \textit{Stability}, and a summarized result is presented in \cref{tab:otherllm}.
It can be observed that in comparison to alternative language models, the GPT-3.5 and GPT-4 models have consistently produced better results for generating the different honeytokens with valid syntaxes. GPT-3.5 has performed consistently well for all the seven honeytoken with valid syntax. Gemini had the most difficulties in the generation of valid syntax, with multiple "x" ratings, indicating that it may not have executed some prompts properly due to underlying model limitations. Upon analyzing the robots.txt file it was observed that GPT-3.5 and GPT-4 performed well across all metrics for the honeytokens. LLaMA, too, showed stable syntax generation across all honeytokens. For the category of ports and services, it is observed that GPT-3.5, GPT-4, and Gemini perform consistently well as compared to LLaMA-2. Similarly, for other honeytokens like invoice files, config files, and log files, Gemini performs the best amongst them across all the metrics. Lastly, for the database honeytoken, only GPT-3.5 could correctly generate the syntax of the database table among all the other \acp{llm}. GPT4 and Gemini could not produce any response to the prompts due to underlying model limitations which did not permit these models to generate database structures. 

It is evident from the tables and figures above that \acp{llm} have effectively generated honeytokens. These documents are enticing to attackers because they contain relevant information and can deceive them. The honeytokens, which have valid syntax, can also be used to fill honeypots that require fake but enticing content. The capability of \acp{llm} to generate honeytokens would allow enterprises to create honeypots for cybersecurity purposes.

\setlength{\tabcolsep}{6pt} %
\renewcommand{\arraystretch}{1}
\begin{table}[ht]
\small %
    \centering
    \caption{Prompts for honeytoken generation evaluated in different LLM, + good, - bad, o neutral, x  not possible to execute. The column labels A-G correspond to IDs in \cref{tab:special_instrcutions}.}
    \begin{tabular}{l|l|c|c|c|c|c|c|c}
    \hline
        \textbf{LLM} & \textbf{Prompt} & \textbf{ A } & \textbf{ B } & \textbf{ C } & \textbf{ D } & \textbf{ E } & \textbf{ F } & \textbf{ G } \\ \hline
        ~ & Syntax &  +  &  + &  +  &  + &  +  &  + &  +   \\ 
        GPT3.5 & Credibility &  +  &  + &  o  &  + &  -  &  + &  -   \\ 
        ~ & Variability &  +  &  + &  o  &  o &  + &  +  &  +  \\ 
        ~ & Stability &  + &  + &  +  &  +  &  + &  +  &  -    \\ 
        ~ & ~ & ~ & ~ & ~ & ~ & ~ & ~ &   \\ 
        ~ & Syntax & + &  + &  + &  + &  o &  + & x  \\ 
        GPT4 & Credibility &  o &  + &  + &  + & o &  + & x  \\ 
        ~ & Variability &  + &  + &  o &  + &  - & o & x  \\ 
        ~ & Stability &  + &  + &  + &  + &  + &  + & x  \\ 
        ~ & ~ & ~ & ~ & ~ & ~ & ~ & ~ &   \\ 
        ~ & Syntax & o &  + &  + &  + &  + &  + &  +   \\ 
        LLaMA & Credibility &  + &  + & - &  + &  + & o &  -   \\ 
        ~ & Variability &  + &  + &  + & o &  o &  -  &  -   \\ 
        ~ & Stability &  + &  + &  + &  +  &  + &  + &  -   \\ 
        ~ & ~ & ~ & ~ & ~ & ~ & ~ & ~ &   \\ 
        ~ & Syntax &  + & x &  + &  + & + & + & x  \\ 
        Gemini & Credibility &  + & x &  + &  o &  +  &  o & x  \\ 
        ~ & Variability & + & x &  +  &  + &  +  &  o & x  \\ 
        ~ & Stability &  + & x &  +  &  + &  + &  + & x  \\\hline
    \end{tabular}
    \label{tab:otherllm}
\end{table}

\section{Discussion}\label{sec:limitations_and_discussion}
Assessing the efficacy of various honeywords remains inherently imperfect, relying heavily on human expertise. Crafting metrics to measure the effectiveness of these honeytokens poses a significant challenge, as there are numerous factors to consider. Despite these limitations, the authors tried to construct two metrics aimed mostly at automating the evaluation of honeytokens. However, developing metrics to prove the credibility and success of a honeytoken can be challenging. An ideal solution would involve deploying honeywords in real-world scenarios and subjecting them to real-world attacks for comparison. While such an approach would offer remarkable insights, the very magnitude of effort required renders it impractical for the potential insights gained. A larger comparison between multiple prompts would be prohibitively costly. 

Nevertheless, the presented findings yield promising results, demonstrating the effectiveness of leveraging \acp{llm} to generate authentic honeytokens. While comprehensive proof remains a subject for future exploration, its necessity may be questioned, given that the results already look convincing enough. 

The overall focus of this work was to highlight the capability of \acp{llm} as generic honeytoken generators. Each category of honeytoken provides enough ground for individual research on the specific kind of honeytoken.
As the results show, even a slight modification of the sentence structure or wording can influence the outcome and performance. This work provides a community hub for exchanging and discussing the best deception honeytoken. \footnote{\url{https://github.com/dfki-in-sec/prompt-collection}}
Regarding honeywords overall it should be noted that honeywords alongside real passwords should not be stored in clear text, which was done in this work only to facilitate evaluation. When an attacker finds a database with hashed honeywords stored alongside real user passwords, the presence of honeywords could slow down a potential brute-force attack. 
This research was heavily focused and oriented on ChatGPT. The evaluation and selection of the best building blocks were performed on ChatGPT. Other models may perform better than ChatGPT. As already mentioned, this decision was made at the beginning of the research, when not even all \acp{llm} that were tested in this work were available in Europe (i.e. Gemini and LlaMA2).

\paragraph{Limitations}

It is important to consider the limitations of \acp{llm} as they significantly impacted the generation of honeytokens. For instance, while generating a complete database with user details, it was observed that some of the \acp{llm} could not generate any database, which could have happened due to various reasons such as complexity, domain knowledge, or policy constraints.
Another limitation based on the token prediction of LLM is that without an external source of randomness, it can't generate random responses, which needs to be considered when generating new data. This is due to the probabilities of token sequences being determined during the training phase, which can lead to repeating patterns in the output. As with most models a temperature parameter can control the variation of the selected token and thus the perceived randomness of the model output. This influenced the results of some tokens, generating repetitive or similar content in some cases. Additionally, it is important to note that \acp{llm} can not provide more recent information than their training phase, which could result in non-evolving content making generated honeytokens more easily detectable through learning their patterns. 
 
Furthermore, the fast and steady development of \acp{llm} and the changing model versions make reproducing the results hard. For instance, the quality of a prompt output can change over time for better or worse \cite{Chen.18.07.2023}. This makes it hard to perform precise research in this domain.

In this work, the authors have presented honeyword generation method which is considerably less detectable than previous methods. However, it must be noted that a direct comparison of these results to earlier results warrants careful consideration due to the slightly smaller scale of the presented approach. An extensive comparison would be more cost-intensive and could be targeted in future work. 

\paragraph{Future Work}
Future research in the field of honeytokens holds significant potential for expansion and refinement. The current research, albeit focused on a limited selection of honeytokens, lays the groundwork for broader implementation. Researchers could explore creating specific types of honeytokens to make them more believable and effective.
Additionally, there's a possibility of developing tools to automatically generate files like PowerPoint or Word documents, streamlining the process of deploying honeytokens. The idea of using honeytokens to set up environments autonomously is also worth investigating further. By integrating honeytokens into existing systems, researchers can gather real-world data to improve their effectiveness and assess their susceptibility to detection. Furthermore, future work could involve fine-tuning \acp{llm} for better honeytoken generation or even creating entire fake companies, including advertising materials, to enhance the effectiveness of honeypots. In addition to the method used in this work, other prompt engineering techniques like few-shot and zero-shot approaches may be considered, or the token amount can be considered as a parameter. 
Fix Idea: For further optimization of the generator prompts, the evaluation metrics could be fed back into an auto-tuning loop in order to validate the effectiveness, and discriminators could be trained. Finally, enhancing the performance and cost efficiency of \acp{llm} may be achieved through fine-tuning, considering the temperature parameter, and adjusting token size. Moreover, exploring smaller transformers can further optimize cost.

\section{Conclusion} \label{sec:conclusion}
In summary, this work examined the possibility of leveraging the generative power of various \acp{llm}, namely GPT3.5, GPT4.0, Gemini, and LLaMA2, for the generation of honeytokens. 
Conversational agents or chatbots available for these \acp{llm} were used. 
The study aimed to generate seven honeytokens, namely robots.txt, configuration files, log files, databases, honeywords, invoice files, and  ports/services specifications. 
As part of the research, experiments were conducted using a varied range of prompts to determine the most effective prompt structures. Prompts were essentially divided into building blocks, each sub-block was then used to build up a complete prompt sequence. A total of 210 prompts were created and tested, and basic statistical analysis was performed to identify the top 20 building blocks. 
Moreover, these prompts were evaluated based on the generation of two specific honeytokens, i.e. honeywords, and robots.txt. The authors devised custom metrics to evaluate the quality of these generated honeytokens.
In addition, flatness was employed as an established metric, with the assessment conducted using Wang et al.'s tool revealing a noteworthy $15.15\%$ success probability of distinguishing attacks, signifying a notable enhancement over Juels and Rivest's respective success rates of $29.29\%$ and $32.62\%$. These findings clearly show that \acp{llm} offers promising generation capabilities. This can be leveraged for various cyber-deception tasks and in the generation of honeytokens. The study reinforces the potential of \ac{llm}-assisted honeytoken generation to develop more sophisticated solutions for cyber deception.

\begin{acks}
This work was supported by the \grantsponsor{BMBF09042021}{German Federal Ministry of Education}{https://www.bmbf.de/bmbf/de/home/home_node.html}
and Research (BMBF) through Open6G-Hub (Grant no.~\grantnum{BMBF09042021}{16KISK003K})
\end{acks}

\bibliographystyle{ACM-Reference-Format}

\bibliography{llm_honeypots}
\appendix

\end{document}